# Entangled Photon Pair Generation in Hybrid Superconductor-Semiconductor Quantum Dot Devices

M. Khoshnegar[1] and A. H. Majedi[1,2,3]

[1]Institute for Quantum Computing, [2]Department of Physics and Astronomy, University of Waterloo, Waterloo, Ontario, Canada N2L 3G1

[3]School of Engineering and Applied Sciences, Harvard University, Cambridge, MA 02138

We investigate the effect of Cooper pair injection in shifting biexciton energy level of low-symmetry ($C_{2v}$) quantum dots (QDs) exhibiting nontrivial fine structure splitting. Coupling QDs to the superconducting coherent state forms extra fine structures by intermixing the ground and biexcitonic states where spectroscopic separation of neutral exciton and biexciton can be diminished, yielding a system to be utilized in time reordering scheme. The separability of exciton and biexciton energy levels is ascribed to the corresponding direct, exchange and correlation energies calculated here through configuration interaction method. We demonstrate the possibility of enhancing photon entanglement concurrence via providing an energy coincidence for biexciton-exciton ($XX \rightarrow X$) and exciton-ground ($X \rightarrow 0$) emissions within the weak coupling regime.

## I. Introduction

On-demand sources of entangled photons are being one of the fundamental building blocks for quantum computing purposes, quantum cryptography, and quantum communication.[1,2] In this context, biexciton-exciton cascade recombination process in semiconductor QDs has already been proposed for generating polarized-entangled photons.[3] However, the intrinsic fine structure splitting particularly exhibited by self-organized QDs, owing to predominant long range electron-hole exchange interaction,[4] degrades the indistinguishability between decay paths and hence entanglement of emitted twin photons. Since this exchange interaction is a direct consequence of lateral anisotropy in electron-hole localizations, it can be vanished via symmetrising the QD carrier confinement,[5] or manipulating strain and hence the built-in piezoelectric fields.[6] Several methods have been exploited earlier to suppress the destructive effect of fine structure splitting (FSS) including perturbation-induced approaches like dc and ac Stark effects,[7] magnetic field Zeeman effect[8] and cavity coupling,[9] or postgrowth techniques such as thermal[10] and laser[11] local annealing, in order to restore group symmetries equal to $C_{4v}$ or $D_{2d}$.[12] An alternative route is growing III-V nanowire-QDs along [001] or [111] crystallographic orientation to principally prevent the atomistic asymmetries distorting QD confinement.[13] Furthermore, the so called "which path" information can be erased through spectral filtering approaches,[14] or the destructive phase developed as a consequence of FSS can be compensated yielding higher fidelities.[15]

One recently proposed method removing any substantial restriction on bright state splitting (BS) is the time reordering scheme,[16] which demands for biexciton binding energy,

$$\delta_{bi} = E^{X_1} + E^{X_2} - E^{XX}, \quad (1)$$

to be zero [see Fig. 1(a)]. In this equation, $\delta_{bi}$ stands for the biexciton binding energy, $E^{X_1}$ and $E^{X_2}$ are the energies of intermediate excitonic states and $E^{XX}$ refers to the energy of biexciton level. Fig. 1(a) shows the diagram of the biexciton-exciton cascade process in a typical QD, where $\delta_{bi}$ is responsible for energy spacing between biexcitonic and excitonic transitions, and $\delta_{bs}$ represents bright state splitting. The ideal level arrangement for time reordering scheme is, however, illustrated in Fig. 1(b), having the biexciton energy level tuned equal to sum of the bright state energies. Accordingly, the first photon in each path is polarized-entangled to the second photon in the other path having identical energies. In this framework, QD structural properties, including material and geometrical parameters as well as strain field can be manipulated in order to erase $\delta_{bi}$ either for binding or anti-binding biexcitons.[17] As a result, the carriers direct and exchange Coulomb interactions slightly vary inside the QD and give rise to a nominal shift in excitonic levels. Practically, this method requires a precise control over QD dimensions and carrier confinement. An alternative solution is to exert lateral electric fields to manage the interplay between single particle Coulomb interactions and shift the biexciton level upward or downward.

Here, we propose applying a trivial perturbation to the original few-particle states by coupling the QD into the coherent state of a low band gap material such as superconductor. From the macroscopic point of view, the proximity effect adjacent to the QD region diminishes the superconductor intrinsic gap $\Delta$ and, therefore, is capable of providing an energy adjustment ranging from QD's BS (< 150 μeV) to biexciton binding energy (few meVs). In contrast to the previously mentioned methods, here the biexcitonic level is displaced while four-fold excitonic fine structure essentially remains isolated of any change.

The paper is organized as follows. In Sec. II we describe the intermixing between the QD excitonic states as a consequence of being coupled to the superconductor coherent state. In Sec. III we briefly explain how the Coulomb interactions in an exemplary QD are calculated. The full description of theoretical modelings can be found in the references addressed within the text. We emphasize that the method of calculation is quite general and could be applied to any QD of the same type. In Sec. IV we explain the possibility of regulating the newly induced fine structure under external voltage bias. In Sec. V the approximate concurrence of entanglement is calculated, and in Sec. VI a summary of our results is provided.

**II. Intermixing of QD states**

Hybrid superconductor-QD devices have already been realised presenting Cooper pair tunneling through QD in three different regimes.[18] Defining $\Gamma$ as the level broadening of the successive tunneling intermediate state, $J$ as the average Coulomb interaction of single particles, and $\Delta$ as the superconducting intrinsic gap, these regimes are categorized as: 1) strong coupling ($\Gamma \gg J$, $\Gamma \gg \Delta$), in which the negligible Coulomb blockade cannot prohibit the Cooper pair tunneling of electrons and holes, thus supercurrent can flow through the device analogous to the single particle current. 2) intermediate coupling, where the energy scales are in the same range, i.e. $\Gamma \sim J$, $\Gamma \sim \Delta$. $\Gamma \sim J$ indicates that coupling is adequate so that considerable suppercurrent appears even when the Fermi level of leads and QD energy levels are off-resonance. Furthermore, $\Gamma \sim \Delta$ ensures that under Fermi level alignment a significant suppercurrent can flow even in the presence of high Coulomb interactions. 3) weak coupling ($\Gamma \ll J$, $\Gamma \ll \Delta$): in the limit of weak coupling, the resonant tunneling of Cooper pairs is predominantly prohibited by Coulomb blockade giving rise to e.g. quasiparticle forth-order co-tunneling mechanisms; however, device still exhibits Josephson junction behaviour with critical current $I_c \sim (2e/\hbar)\Gamma^2/\Delta$.[19] Nevertheless, suppercurrent can transfer through the QD via higher-order quantum coherent tunneling processes.[20] The effective hamiltonian of the superconductor-coupled QD for the *isolated* electrons (neglecting the presence of hole particles) reads[21, 22]

$$\widehat{\mathcal{H}}_e = \sum_{\sigma_e} E_e \hat{c}^\dagger_{\sigma_e} \hat{c}_{\sigma_e} + J_{ee}\hat{n}_\uparrow \hat{n}_\downarrow + \widetilde{\Delta}_e \hat{c}^\dagger_\uparrow \hat{c}^\dagger_\downarrow + \widetilde{\Delta}^*_e \hat{c}_\uparrow \hat{c}_\downarrow \qquad (2)$$

where $\hat{c}^\dagger_{\sigma_e}$ ($\hat{c}_{\sigma_e}$) creates (annihilates) an electron possessing spin $\sigma_e \in \{1/2, -1/2\}$, and $\hat{n}_{\sigma_e}$ strands for the number operator of the same particle. $E_e$ and $J_{ee}$ are the single electron kinetic energy, and electron-electron on-site Coulomb repulsion, respectively. The collective motion of Cooper pairs experiences a suppression of superconductivity when approaching the QD, known as proximity effect, where the intrinsic gap begins to diminish and an effective superconducting gap $\widetilde{\Delta}_e$ is defined. $\widetilde{\Delta}_e$ depends on the Coulombic interactions and level broadening, thus need to be renormalized when QD energy levels change.[20] An exactly analogous hamiltonian governs the dynamics of *isolated* holes, and Eq. (2) is valid when energies are redefined for hole particles, and $\hat{c}^\dagger_{\sigma_e}$, $\hat{c}_{\sigma_e}$ and $\hat{n}_{\sigma_e}$ are replaced with $\hat{h}^\dagger_{\sigma_h}$, $\hat{h}_{\sigma_h}$ and $\hat{n}_{\sigma_h}$. The only difference lies in the pseudo-spin of holes: valence subbands of III-V materials in zinc-blende or wurtzite phase are categorized into three families near $\Gamma_8$ or $\Gamma_7$ valleys, including heavy-hole (HH), light hole (LH) and spin orbit split-off (SO). Depending on the position of QD level in the energy space it has contribution from all these bands with different weights as a consequence of band mixing. However, the predominant contribution to the valence band (VB) ground state (the topmost energy level) comes from HH band, where the $z$-projection of total angular momentum, or equivalently the hole's pseudo-spin, is $\sigma_h \in \{3/2, -3/2\}$. The second contribution associated with the closest LH band is trivial especially when the QD height is small and vertical confinement becomes strong.

In order to determine the realistic energy levels of the coupled electron-hole system, their direct and exchange Coulomb interactions must be considered. Diagonalizing the hamiltonian in Eq. (2) leads to one unperturbed doublet having energies $\xi_{\uparrow D} = \xi_{\downarrow D} = \xi_D$ (hereinafter, energies are indicated with respect to the superconductor chemical potential, i.e. $\xi_D = E_e - \mu^e$), and two singlets with energies $\xi_{S_0, S_1} = \xi_D + J_{ee}/2 \pm [(\xi_D + J_{ee}/2)^2 + |\widetilde{\Delta}_e|^2]^{1/2}$ being mixed of the ground and *one-pair excited* states[21]:

$$|e_{0;e}\rangle_S = -e^{-i\varphi_e}|u_e||G_e\rangle + |v_e||\Theta_e\rangle, \qquad (3.a)$$

$$|e_{1;e}\rangle_S = e^{-i\varphi_e}|v_e||G_e\rangle + |u_e||\Theta_e\rangle \qquad (3.b)$$

where $|G_e\rangle$ corresponds to BCS ground state, and $|\Theta_e\rangle$ represents the situation where one Cooper pair of electrons is transferred into the QD ($|\Theta_e\rangle = \hat{c}_\uparrow^\dagger \hat{c}_\downarrow^\dagger |G_e\rangle$); $v_e$ and $u_e$ are coherence factors denoting the probability of a state being occupied by Cooper pairs $|v_e|^2$ or quasi-particles $|u_e|^2$.[23] $\varphi_e$ stands for the superconducting condensate phase and subscript "$S$" ascribes these eigenfunctions to singlet mixed states. Doublet is the *bare* QD single particle state experiencing no energy variation under above circumstance; however, it gives rise to the possible odd parity cycles and is of importance when trionic optical transitions, $X^+$ and $X^-$ are included in the model.[21,24] These odd cycles seem to be a principal issue in weak coupling limit where number of quasi-particles entering the QD is sizable in comparison with Cooper pairs: in the case that quasi-particle tunnelling is the superior injection mechanism, negatively $X^-$ or positively $X^+$ charged excitons are expected to inevitably contribute to the spectrum although their sharp-line emissions can be filtered out. Nevertheless, to simplify the model we neglect odd cycles in this work. A similar treatment is applicable to the hole side again by replacing $|G_h\rangle$ and $|\Theta_h\rangle$ in (2), where $|\Theta_h\rangle = \hat{h}_\uparrow^\dagger \hat{h}_\downarrow^\dagger |G_h\rangle$. Expanding the total hamiltonian

$$\widehat{\mathcal{H}}_{tot} = \widehat{\mathcal{H}}_e + \widehat{\mathcal{H}}_h + \sum_{\sigma_e \sigma_h} J^{d,ex,corr} \hat{n}_{\sigma_e} \hat{n}_{\sigma_h} \qquad (4)$$

on mixed subspace $\{|e\rangle\otimes|h\rangle\} = \{|e_0, h_0\rangle, |e_1, h_0\rangle, |e_0, h_1\rangle, |e_1, h_1\rangle\}$, i.e. $\mathcal{H}_{mn} = \langle h, e|_n \mathcal{H}_{tot}|e, h\rangle_m$, induces four-fold fine structures in the vicinity of both ground and biexcitonic states. This induced fine structure is schematically illustrated in Fig. 1(c). In the expression above, $J^d$, $J^{ex}$ and $J^{corr}$ are the QD ground state direct, exchange and correlation energies. The energy regulation of extra sub-levels then depends on two parameters: chemical potential of the leads $\mu_l$ and the tunnelling rate already preserved in $\widetilde{\Delta}$ (we assume that the superconducting effective gap is equal for both electron and hole sides). It is convenient to go to the number state representation by performing a unitary transformation $\mathcal{T}$ with matrix elements depending on $\mu_l$, $\widetilde{\Delta}$ and superconducting phase $\varphi_{e,h}$, i.e. $\{|G_e, G_h\rangle, |\Theta_e, G_h\rangle, |G_e, \Theta_h\rangle, |\Theta_e, \Theta_h\rangle\} = \mathcal{T}(v_{e,h}, u_{e,h}, \phi_{e,h}) \{|e\rangle\otimes|h\rangle\}$. As the QD bright states are essentially fixed, $|G_e, G_h\rangle$ and $|\Theta_e, \Theta_h\rangle$ energy levels can be finely tuned to effectively remove the biexciton binding energy. Henceforth, we label the energy of the four induced states as follows: $E_0 = E_{|e_0,h_0\rangle}$, $E_1 = E_{|e_1,h_0\rangle}$, $E_2 = E_{|e_0,h_1\rangle}$ and $E_3 = E_{|e_1,h_1\rangle}$.

**III. QD few-particle interactions**

Here we explain the proposed method by giving a relevant specific example. The approach of calculations given here, consisting of *k.p* model and configuration interaction (CI) method, is however quite general for III-V materials in zinc-blende phase.[25] In the case where QD emitter is made of wurtzite-structured material the appropriate single particle *k.p* hamiltonian must be replaced.[25] Necessity of calculating the on-site Coulomb interactions relies on these facts: first, it confirms the ordering of $\Gamma$, $J$ and $\Delta$, thus the regime of device operation. Secondly, it is a commonly used experimental method to tune the QD energy levels and also its electron and hole Coulomb interactions by applying an appropriate gate voltage. This gate voltage induces an electric field mostly along the lateral directions of QD, thus separates electron and hole probability densities leading to lower electron-hole interactions, together with a reduction in excitonic oscillator strength which is not favourable in photon emission applications. However, as long as the oscillator strength is not drastically suppressed, gate manipulation is a practical solution for tuning the Coulombic interactions of QD. In the following we show that for conventional QDs working as single photon sources in the infrared wavelength range, the Coulombic interactions are adequately large that energy ordering of $\Gamma$, $J$ and $\Delta$ dictates the device operation to be in the weak coupling regime.

Our setup consists of a typical single self-assembled InAs/GaAs quantum dot grown along [001] direction on top of a 2 ML InAs wetting layer capped by GaAs as the local barrier and connected laterally to superconducting electrodes.[19] The biexciton binding energy which is defined with respect to the recombination energy of the exciton ground state, $\delta_{bi} = \hbar\omega_{|X\rangle_D \to |G\rangle_D} - \hbar\omega_{|XX\rangle_D \to |X\rangle_D}$, reads[26]

$$\delta_{bi} = -2J_{eh}^d - J_{ee}^d - J_{hh}^d + 2J_{|X\rangle_D}^{corr} - J_{|XX\rangle_D}^{corr} + 2J_{|X\rangle_D}^{ex} - J_{|XX\rangle_D}^{ex} \qquad (5)$$

where subscript "D" stands for the excitonic levels of *bare* dot. The direct terms, $J_{ee}^d$, $J_{hh}^d$ and $J_{eh}^d$, along with exchange terms, $J_{|X\rangle_D}^{ex}$ and $J_{|XX\rangle_D}^{ex}$ contributing to the expression above are primarily determined by taking only wavefunctions of QD ground state into account: the single particle wavefunctions of carriers trapped inside QD is composed of two components; the Bloch spinor and its associated envelope function. For instance, in the simplest case, the total wavefunction of an electron in the S-orbital of conduction band (CB) can be represented by $\psi_{\sigma_{e_0}}^e = \phi_0^e |S_{CB}; \sigma_{e_0}\rangle$ and similarly the total wavefunction of a hole in the P-orbital of valence heavy hole band is given by $\psi_{\sigma_{h_0}}^h = \phi_0^h |P_{HH}; \sigma_{h_0}\rangle$. However, there is a coupling coefficient between the conduction and valence wavefunctions, included in the *k.p* model, originating from the QD asymmetries. These coupling terms are also responsible for the non-zero long range exchange interaction which gives rise to the splitting of bright states. As long as QD confinement maintains laterally symmetric, the coupling remains trivial. On the other hand, as mentioned above, the QD valence ground state is usually an admixed of HH and LH bands commonly with larger weight from the HH part. The mixing coefficient between these two subbands depends on the QD geometry, especially its vertical anisotropy: the large amount of anisotropy in the vertical orientation (growth direction) of QD leads to higher mixing orders between HH and LH. Consequently, nearly flat QDs keep the VB ground state relatively HH-type in character. Fig. 2(c) shows the density of coulomb interaction matrix elements[4]

$$\rho_{p_0 q_0, \sigma\acute{\sigma}} = \psi_{\acute{\sigma}_{q_0}}^{q_0}{}^* \psi_{\acute{\sigma}_{p_0}}^{p_0}{}^* C(\mathbf{r}_1, \mathbf{r}_2) \psi_{\sigma_{p_0}}^{p_0} \psi_{\sigma_{q_0}}^{q_0} \qquad (6.a)$$

$$\mathcal{H}_{p_0 q_0; \sigma\acute{\sigma}}^{|X\rangle_D} = \left\langle \psi_{\acute{\sigma}_{p_0}}^{p_0} \psi_{\acute{\sigma}_{q_0}}^{q_0} \middle| C(\mathbf{r}_1, \mathbf{r}_2) \middle| \psi_{\sigma_{p_0}}^{p_0} \psi_{\sigma_{q_0}}^{q_0} \right\rangle \qquad (6.b)$$

where $C(\mathbf{r}_1, \mathbf{r}_2) = e^2/4\pi\epsilon(\mathbf{r}_1, \mathbf{r}_2)|\mathbf{r}_1 - \mathbf{r}_2|$. Here, $p_0, q_0 \in \{e_0, h_0\}$ represent particle labels dwelling in the ground states, and $\epsilon$ denotes the static dielectric constant. The corresponding direct interactions can be then evaluated as $J_{p_0,q_0}^d = \langle \psi_{\acute{\sigma}}^{p_0} \psi_{\acute{\sigma}}^{q_0} | C | \psi_{\sigma}^{p_0} \psi_{\sigma}^{q_0} \rangle$, where $\sigma\acute{\sigma} = \uparrow\downarrow$ if $p_0 = q_0$, else $\sigma\acute{\sigma} = \uparrow\uparrow$. Analogously the ground state exchange terms $J_{p_0,q_0}^{ex}$, which originate the bright and dark states splittings, are calculated replacing $|\psi_\sigma^{p_0}(\mathbf{r}_1)\psi_\sigma^{q_0}(\mathbf{r}_2)\rangle$ by $|\psi_\sigma^{p_0}(\mathbf{r}_2)\psi_\sigma^{q_0}(\mathbf{r}_1)\rangle$ in $J_{p_0,q_0}^d$. Above single band treatment, however, only covers the Fermi correlations. To account for the Coulomb correlations arising in many-body systems, higher energy levels shall be regarded: Eight-band *k.p* hamiltonian[25] was employed to solve for the single particle wavefunctions of a 25 nm-long in base ($b_D$= 25 nm), 5 nm-high ($h_D$= 5 nm), zinc-blende InAs/GaAs truncated pyramidal QD. The vertical aspect ratio of $h_D/b_D$= 0.2 provides the required separability between HH and LH bands, hence, the pseudo-spin of holes in VB ground state is semi-purely ± 3/2. Besides, our calculations show that for this vertical aspect ratio, the electron-electron and hole-hole repulsive interactions are both comparable to electron-hole attractive interaction, yielding less biexciton binding energy. The required material parameters, including *k.p* parameters and elasticity constants, are taken from references 13 and 23. Provided the single particle states by *k.p* method, CI approach was then utilized to construct the true wave function of few particle multiexcitons and calculate the direct and exchange Coulomb interactions along with the correlation energies:[26] 10 hole and 8 electron subbands were included to build up the configuration set. The calculated exciton and biexciton energy levels are depicted in Fig. 2(a). The negligible intrinsic BS (< 2 μeV) obtained here is a consequence of laterally symmetric and vertically flat QD.[13] Particularly, the QD flatness suppresses the role of strain-induced piezoelectric polarization,[27] which plays a destructive role on the $C_{4v}$ symmetry of electron and hole envelope functions and, hence, the degeneracy of QD bright states. Although the QD studied here is potentially an ideal source of entangled photons due to its small BS, the nondeterministic growth process of QDs giving rise to possible geometrical imperfections, puts no guarantee on their lateral symmetry in practice.[5] Therefore, in the present work we focus on optimally removing the biexciton binding energy, making cascade levels appropriate for time reordering measurements, while the conclusions given in the following are legitimate also for large BS QDs.

From CI calculations, we exploited $J_{eh} = J_{eh}^d = -12.54$ meV, $J_{ee} = J_{ee}^d = 12.97$ meV and $J_{hh} = J_{hh}^d = 13.2$ meV much larger than the intrinsic gap $\Delta$ of a typical low $T_c$ superconductor like niobium ($\Delta_{Nb} \approx 1.5$ meV at T = 0). This indicates that: 1) in QDs having almost vertical symmetry, where the electron and hole probability densities are similarly localized in space, direct Coulomb interactions are comparable and relatively cancel out each other in Eq. (5). The summation of the exchange ($\sum_{|X\rangle_D,|XX\rangle_D} J^{ex} \leq \Delta$) and correlation ($\sum_{|X\rangle_D,|XX\rangle_D} J^{corr}$) energies then significantly participate in $\delta_{bi}$ determination. 2) having $\widetilde{\Delta}_{e,h} = \Gamma_{e,h}\exp(i\varphi_{e,h})/2$,[28] the condition $J^d \gg \Delta$ ensures that device operates in the weak coupling regime and $|G_e, G_h\rangle$ is the equilibrium quasi-state since the probability of being in quasi-state $|\Theta_e, \Theta_h\rangle$ is given by $|\widetilde{\Delta}_{e,h}|^4/J^{d^4} \ll 1$ in second order perturbation.[22] Hereinafter, we assume that the chances of electrons passing into and escaping from the QD states are equal, i.e. $\Gamma = \Gamma_e = \Gamma_h$. This presumption is valid until device is symmetric and simplifies the model without loss of generality.

**IV. Mixed states fine tuning**

In order to activate the recombination process, quasi state $|\Theta_e, \Theta_h\rangle$ must be pumped up by imposing a bias voltage into the leads to separate their chemical potentials, $\mu_l^e$ and $\mu_l^h$, close to the QD excitonic band gap. Meanwhile, by adjusting the detuning $\Delta\varepsilon_{e,h}$, defined as the energy spacing between chemical potentials and single particle energy levels, $E_e$ and $E_h$, $|G_e, G_h\rangle$ and $|\Theta_e, \Theta_h\rangle$ quasi-states tend to a selected degeneracy point.[22] Sequential injection of Cooper pairs of electrons (holes) then fills up the intermediate $|\Theta_e, G_h\rangle$ ($|G_e, \Theta_h\rangle$) and finally $|\Theta_e, \Theta_h\rangle$ quasi-states. Subsequently, $|\Theta_e, \Theta_h\rangle$ decays radiatively into the ground state $|G_e, G_h\rangle$ through intermediate excitonic bright states and closes the cycle by emitting polarized photons.

One technique to directly reduce $|G_e, G_h\rangle$ and $|\Theta_e, \Theta_h\rangle$ energy level splitting $\delta_s$ is controlling over chemical potential detuning, under the situation where $J_{ee}, J_{hh}, J_{eh} \gg \Delta$ and $\sum_{|X\rangle_D, |XX\rangle_D} J^{corr} + J^{ex} \lessgtr \Delta$. The latter condition is evidently determined by the type of superconductor and also the geometry of QD. For instance, the calculations given by A. Schliwa et al. in Ref. 26 demonstrate that the biexciton correlation energy can even reach half of the direct interaction energies for large vertical aspect ratios. These calculations also confirm that both exciton and biexciton correlation energies are minimized for the vertical aspect ratio we have chosen in our example (0.2). Our calculations exhibit an acceptable consistency with their results in terms of the evolution of exciton and biexciton correlation energies versus vertical aspect ratio. In our case, the correlation terms were estimated $< 2$ meV, depending on the number of basis regarded for the configuration space.

We spanned detuning over the energy range $-J_{eh}^{XX}$ to $J_{eh}^{XX}$, the largest energy scale among Coulomb interactions ($J_{eh}^{XX} = 4J_{eh}$, since the total electron-hole interaction for the biexciton is established between the four constituting single particles of two excitons, and the change of spin configurations between these excitons does not affect the direct interactions), and plotted the eigenfunctions of hybrid system fine structure. The informative part of the plot is shown in Fig. 3(a) and (b): the approximate hopping energy for electron (hole) is presumed to be $J_{ee}^d/2$ ($J_{hh}^d/2$) without loss of generality. Then by setting $\Delta\varepsilon_{e0} = J_{eh}^{XX}/2 + J_{ee}/2$ and $\Delta\varepsilon_{h0} = -J_{eh}^{XX}/2 + J_{hh}/2$ the quasi-states $|G_e, G_h\rangle$, $|\Theta_e, G_h\rangle$ and $|\Theta_e, \Theta_h\rangle$, or their equivalent mixed states, reside nearby at one specific degeneracy point [see anticrossing in Fig. 3(a)]. At the immediate vicinity of anticrossing, which never vanishes as long as the superconducting gap exists, mixed states have contribution from all above quasi-states. Along the skew narrow region, the induced ground $|G\rangle$ and biexciton $|XX\rangle$ states can be estimated through $|G\rangle \approx \alpha_1 |G_e, G_h\rangle + \beta_1 |\Theta_e, \Theta_h\rangle$ and $|XX\rangle \approx \alpha_2 |G_e, G_h\rangle + \beta_2 |\Theta_e, \Theta_h\rangle$. Instead, $|G\rangle$ mixed state becomes a combination of $|G_e, G_h\rangle$ and $|\Theta_e, G_h\rangle$ adjacent to the vertical narrow region, and finally turns into a pure state elsewhere. Fig. 3(a) and (b) depict $\alpha_i$ and $\beta_i$, $i = \{1,2\}$, coefficients of the ground and biexciton states for $\Gamma = 0.5\Delta$ and $1.5\Delta$, where $\Delta$ is assumed to be equal to the superconducting gap of niobium at T= 0, $\Delta \approx \Delta_{Nb} = 1.52$ meV. Changing the level broadening $\Gamma$ over the range ~$0.1\Delta$ to ~$3\Delta$ reproduces the same patterns shown here, but $\Gamma$ value always regulates the gap opened at anticrossing. Moving along the skew region, which is shown to be the approximate degeneracy area of $|G_e, G_h\rangle$ and $|\Theta_e, \Theta_h\rangle$ quasi-states in Fig. 3(c), toward anticrossing, $\beta_i$ decreases giving rise to a transition into $|G_e, G_h\rangle$ state. For initiating the sequential carrier-tunneling photon-generation cycle, we need to set the operational point very close to the anticrossing where $|G_e, G_h\rangle$, $|\Theta_e, \Theta_h\rangle$ and their intermediate level $|\Theta_e, G_h\rangle$ come to degeneracy. We name the respective detunings leading to this degeneracy as $\Delta\varepsilon_{e0}$ and $\Delta\varepsilon_{h0}$. Such an initialization is clarified in Fig. 3(c) for $\Gamma = 0.5\Delta_{Al}$, where we replaced niobium with aluminum in our model to acquire higher energy precision appropriate for fine-tuning the splitting $\delta_s$ ($\Delta_{Al} \sim 165$ µeV $\sim 0.1\Delta_{Nb}$ at T = 0): the bold blue area shows the regions where the difference between ground and biexciton energy levels, $E_3$ and $E_0$, is less than $\Delta_{Al}$. Analogously the pale blue area specifies where the ground and $|\Theta_e, G_h\rangle$ intermediate quasi-states separate less than $\Delta_{Al}$.

Although $\mu_l^e$ and $\mu_l^h$ could be modified to shift $\delta_s = E_3 - E_0$ over a relatively large range, i.e. at least in the order of $\Delta$, they might not be able to explicitly minimize this splitting for a certain value of $\widetilde{\Delta}$. We set detunings $\Delta\varepsilon_e$ and $\Delta\varepsilon_h$ on the skew curve $\Gamma = 0.5\Delta_{Al}$, about $5\widetilde{\Delta} = 1.25\Delta_{Al}$ away from the exact degeneracy point yielding to ~15 µeV splitting. According to the expected BS exhibited by self assembled QDs (20–100 µeV),[8,29] such a splitting energy implies no advantage of time reordering scheme over regular polarized-entangled photon generation method.[3] This brings us to the conclusion that although chemical potentials can provide a wide sweep range for biexciton binding energy, an effective suppression in $\delta_s$, and hence required energy resolution for entanglement purposes, might not necessarily occur.

**V. Tunneling ratio and Concurrence**

The tunneling ratio of Cooper pairs suggests an extra degree of freedom to the above system since it directly determines the level broadening $\Gamma(E) = \sum_m |q_m|^2 \delta(E_m - E)$,[20] where $q_m$ is the transmission probability to the $m$th QD energy level. The underlying concept would be then similar to pushing the system to an alternative *skew region* near the fixed anticrossing point $(\Delta\varepsilon_{e0}, \Delta\varepsilon_{h0})$. Fig. 3(d) and (e) describe how $\delta_s$ variations might be connected to $\Gamma$ in the weak coupling limit (if we call the level broadening at which energy spacing between $E_3$ and $E_0$ is minimized as critical $\Gamma$ labeled by $\Gamma_c$, then the conditions $\Gamma_c \approx 0.2\Delta$, $0.25\Delta$ and $\Gamma_c \ll J_{ee}^d, J_{hh}^d, J_{eh}^d$ confirm the operation in weak coupling regime), and manifests capability of $\delta_s$ to be reduced below radiative linewidth of exciton in typical self-assembled QDs (e.g. ~ 4 µeV).[29,30] In these two plots, we first set the chemical potentials on P$_1$ and P$_2$, then minimize $\delta_s$ at some point by smoothly fluctuating one of the chemical potentials, here $\mu_l^e$, around the initial point and sweeping over $\Gamma$. This mostly happens moving deeper into the weak coupling regime, however the least probability of pair tunneling must be consistently satisfied.

It is noteworthy to mention that other than the amount of $\delta_s$, the distinguishability of photons generated in each decay path also relies on the *bare* QD excitonic linewidths together with superconducting coherence factors, i.e. ($\hbar=1$)

$$\mathcal{R}_{|XX\rangle\to|X_\lambda\rangle} = \gamma_{\text{ph}}|\beta_2(u_{e,h},v_{e,h})|^2 \gamma_{|\Theta_e,\Theta_h\rangle\to|X_\lambda\rangle} \qquad (7.\text{a})$$

$$\mathcal{R}_{|X_\lambda\rangle\to|G\rangle} = \gamma_{\text{ph}}|\alpha_1(u_{e,h},v_{e,h})|^2 \gamma_{|X_\lambda\rangle\to|G_e,G_h\rangle} \qquad (7.\text{b})$$

where $|X_\lambda\rangle$ is the excitonic intermediate state with polarization $\lambda = \{H,V\}$, $\gamma_{\text{ph}}$ denotes the photon linewidth, and the excitonic transition rate reads $\gamma_{|i\rangle\to|f\rangle} = |\langle f|\hat{\mathcal{H}}_{\text{em}}|i\rangle|^2$, having

$$\hat{\mathcal{H}}_{\text{em}} = \Sigma_{S;p;\lambda} g^S_{p\lambda} \hat{a}^\dagger_{p\lambda} \hat{b}_\lambda + \text{H. c.}, \qquad (8)$$

In above equation, $S = \{|\Theta_e,\Theta_h\rangle \to |X_{H,V}\rangle, |X_{H,V}\rangle \to |G_e,G_h\rangle\}$, $\hat{b}_H = 1/\sqrt{2}(\hat{h}_\downarrow \hat{c}_\uparrow + \hat{h}_\uparrow \hat{c}_\downarrow)$, $\hat{b}_V = i/\sqrt{2}(\hat{h}_\downarrow \hat{c}_\uparrow - \hat{h}_\uparrow \hat{c}_\downarrow)$, $\hat{a}^\dagger_p$ creates a photon in $p$th optical mode, and $g^S_\lambda$ incorporates the oscillator strength in each excitonic transition. According to Eqs. (7.a) and (7.b), since the individual linewidths of cross generated photons are affected by $|\alpha_1|, |\beta_2|$ factors, the degree of entanglement can be either ruined or improved in our hybrid system. Assuming $\mathcal{R}_{XX} = \mathcal{R}_{|XX\rangle\to|X_V\rangle} = \mathcal{R}_{|XX\rangle\to|X_H\rangle}$ and $\mathcal{R}_X = \mathcal{R}_{|X_H\rangle\to|G\rangle} = \mathcal{R}_{|X_V\rangle\to|G\rangle}$, one can evaluate the concurrence for the generated states of photons as a measure of entanglement [16]:

$$C = \frac{4\zeta_{XX}\zeta_X}{\pi^2}\left|\int \frac{W_{opt}(\omega_m,\omega_n)}{|\omega_m+\omega_n-\omega_{XX}-i\zeta_{XX}|^2 (\omega_m-\omega_{|X_H\rangle}+i\zeta_X)(\omega_m-\omega_{|X_V\rangle}+i\zeta_X)} d\omega_m d\omega_n\right| \qquad (9)$$

where $\omega_{XX}$ and $\omega_{|X_{\lambda=H,V}\rangle}$ are the biexciton and exciton frequencies, and the spectral half widths of exciton and biexciton are represented by $\zeta_X$ and $\zeta_{XX}$: $\zeta_X = \mathcal{R}_X/2\gamma_{\text{ph}}$ and $\zeta_{XX} = \mathcal{R}_{XX}/2\gamma_{\text{ph}}$. One can simply manipulate the additional phase $W_{opt}$ above in order to enhance the concurrence. The best choice of $W_{opt}$ is still under debate (see papers in Ref. 16), however in the simplest case one can introduce an optical delay $\tau_0$ to add a linear phase, i.e. $W_{opt} = \exp[i(\omega_m - \omega_n)\tau_0]$. Under the condition where color coincidence between the biexciton and exciton is prepared, $\tau_0$ can be optimized to provide the maximum concurrence. P. Pathak and S. Hughs have shown that this time delay only depends on the exciton and biexciton spectral half widths: $\tau_0 = \ln(1 + \zeta_{XX}/2\zeta_X)/\zeta_{XX}$.[16] However, out of the color coincidence condition this time delay must be altered depending on the biexciton binding energy in order to optimize the concurrence. Generally, two parameters determine the amount of concurrence in the time reordering scheme applied to biexciton cascade process: 1) $\delta_b/\zeta_X = (\omega_{XX} - \omega_{|X_H\rangle} - \omega_{|X_V\rangle})/\zeta_X$, where $\delta_b$ is the energy separation of the existing biexciton level and the ideal one when the color coincidence of excitonic and biexcitonic transition occurs. In our model, $\delta_s$ is a good measure to represent this energy spacing. 2) The ratio between the exciton and biexciton lifetimes, or equivalently their spectral linewidths $\zeta_{XX}/\zeta_X$. Going back to Eqs. (7.a) and (7.b), this ratio relies upon $\alpha_1$ and $\beta_2$ in addition to bare QD transition rates,

$\gamma_{|\Theta_e,\Theta_h\rangle \to |X_{H,V}\rangle}$ and $\gamma_{|X_{H,V}\rangle \to |G_e,G_h\rangle}$. Near anticrossing shown in Fig. 3(a), where $|\alpha_1|, |\beta_2| < 1$, concurrence $C$ comprises $\delta_s$, $\alpha_1$ and $\beta_2$ information as a consequence of intermixing. This reflects that by tuning $\Delta\varepsilon_e$, $\Delta\varepsilon_h$ and $\Gamma$ in order to minimize $\delta_s$, the potentially maximum concurrence might not necessarily be achieved. However, by controlling over $W_{opt}$ reaching the optimum limit is feasible when realistic values of excitonic broadenings, $\gamma_{|\Theta_e,\Theta_h\rangle \to |X_\lambda\rangle}$ and $\gamma_{|X_\lambda\rangle \to |G_e,G_h\rangle}$, are known.

We examined the evolution of concurrence $C$ given in Eq. (9) versus $\Gamma$ in the same sense accomplished for $\delta_s$ [see Fig. 4(a)]. We biased the model exactly at $P_2$ [see Fig. 3(c)] by setting the appropriate chemical potentials $\mu_l^e$ and $\mu_l^h$, then changed the level broadening $\Gamma$ over a relatively large scale (curve $C_1$). For smaller values of $\Gamma$, the concurrence is merely a function of $\delta_s/\zeta_X$, since both $|\alpha_1|$ and $|\beta_2|$ are almost equal [see Fig. 4(b)] and the $\zeta_{XX}/\zeta_X$ ratio remains constant. In contrast, for larger values of $\Gamma$, $|\alpha_1|^2/|\beta_2|^2$ is increased as can be deduced from Fig 4(b), giving rise to smaller $\zeta_{XX}/\zeta_X$ ratios and a local rise in concurrence. We also changed $\mu_l^e$ by 2 μeV steps toward anticrossing to show how concurrence might be improved for smaller $\delta_s$ values; see curves $C_2$ and $C_3$ in Fig 4(a). In Ref. 16, Avron *et al*. demonstrated that for a fixed amount of $\zeta_{XX}/\zeta_X$, the concurrence is enhanced when the $\delta_b/\zeta_X$ approaches zero. However, in our model both $\delta_s$ and $\zeta_X$ undergo a significant change moving deeper into the weak coupling regime (e.g. $\Gamma < 0.4\Delta$), where the interplay between their variations leads to a local maximum of concurrence next to the anticrossing.

We note that here the cascade process is presumed to be isolated of cross-dephasing between $|X_H\rangle$ and $|X_V\rangle$ which indeed lowers the concurrence in practice, and is a fundamental issue when intermediate exciton states are not identical nor symmetrically coupled.[30] Notice also that throughout the analysis we restricted the operation regime into the weak coupling limit where the hybridization factor $|\widetilde{\Delta}|$, as the key property of setup, was represented by $\Gamma/2$, whereas in practice the relative measure of $\widetilde{\Delta}$ and $\Gamma$ is not fully restrained.[21,22] Reminding the fact that the intermixing phenomenon explicitly depends on the effective superconducting gap, there is a possibility to reduce it for the sake of energy resolution without suppressing tunneling probability. A simple procedure for dynamical modification of effective gap might be the application of a small magnetic field which leaves the QD features unaffected. Another option would be exploiting a back-gate to manage the charging and hence Coulomb interactions inside QD, and eventually impose required changes on $\Gamma$ and $\widetilde{\Delta}$.[31] However in the latter case, losing a part of oscillator strength seems inevitable.

**VI. Summary**

In conclusion, by providing a relevant example we studied the applicability of superconductor-coupled QDs in enhancing the degree of entanglement via suppressing the biexciton binding energy under time reordering scheme. This method allows for tuning biexciton binding energy over a relatively large energy range, in contrast to the setups utilizing lateral strain[17] or local electric filed,[32] by forming extra fine structures near ground and biexcitonic energy levels, and hence results in the observation of well-defined entangled photon pair state. The reason is that here the new energy levels commute somehow independent of the QD confinement and its original fine structure, but rather are linked to the characteristics of superconducting contacts such as $\Delta$ and $\Gamma$. We believe that appropriate contacting of large BS QDs, rendering weak coupling, can optimize the concurrence even in a laser- or cavity-free setup and without requiring any postgrowth manipulation of QDs.

## Acknowledgement

We would thank A. J. Salim and M. Ansari from Quantum Device group at IQC for their fruitful discussions on Cooper pair tunneling through semiconductor QDs.

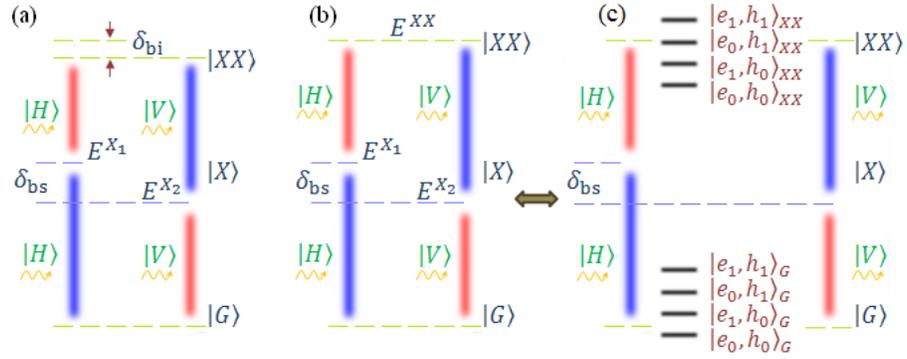

FIG. 1 (a) Biexciton-exciton cascade recombination process emitting two photons with identical rectilinear polarization, but dissimilar energies within each path ($|H\rangle$ and $|V\rangle$ stand for photon polarization states in linear representation). $\delta_{bs}$ and $\delta_{bi}$ denote the bright state splitting and biexciton binding energy of QD, respectively. (b) In time reordering scheme where $\delta_{bi}$ is removed, photons having orthogonal (for circular polarization: parallel) polarization states are entangled so that properly erase the "which path" information. (c) Cascade recombination in perturbed QD coupled to superconducting leads. QD ground state biexcitonic singlet is split into four levels as a result of being coupled to the BCS coherent state. Depending on the Cooper pair tunnelling ratio through QD island, four levels redshift or blueshift and become closely degenerate under special circumstances, eventually developing a new tunable fine structure.

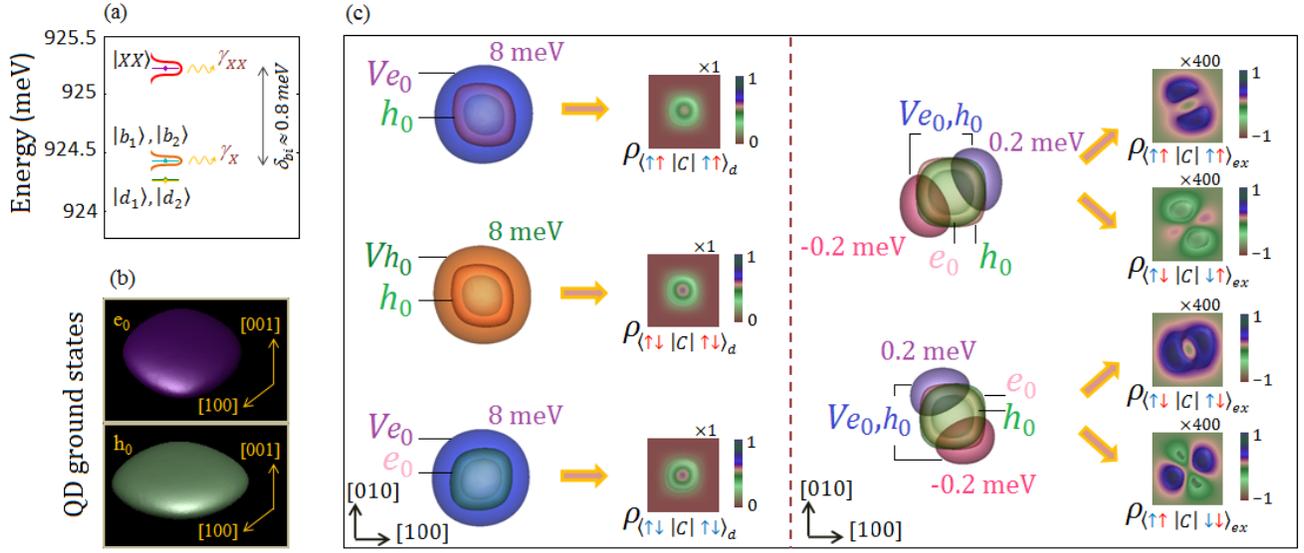

FIG. 2 (a) Biexciton $|XX\rangle$ and exciton, bright $|b_1\rangle, |b_2\rangle$ and dark $|d_1\rangle, |d_2\rangle$ singlets, energy levels of a 25 nm-long, 5 nm-high [001]-oriented InAs/GaAs truncated pyramidal QD on a 2 ML InAs wetting layer before being coupled to superconducting leads. Biexciton is anti-binding with $\delta_{bi} \approx -0.8$ meV. (b) Electron and hole ground state probability densities, $e_0$ and $h_0$, slightly extended along [110] and [1$\bar{1}$0] directions as a consequence of the existing piezoelectric polarization. QD flatness minimizes heavy hole (HH) and light hole (LH) intermixing giving rise to significant isolation of the first HH band. As a result, the most contribution to $h_0$ probability density is attributed to HH envelope function. (c) Left panel: direct coulomb interactions, $J_{ee}$, $J_{hh}$ and $J_{eh}$, can be obtained by locating the charge density of one single particle in the mean-field potential caused by the other single particle. $V_{e_0}$ and $V_{h_0}$ are the electron- and hole-induced potentials, and $e_0$ and $h_0$ label electron and hole probability densities. The normalized density of CI matrix elements, $\rho_{p_0 q_0, \sigma\acute{\sigma}}$, for the ground state direct interaction terms are plotted on (001) plane 1 nm above the QD base. Each $\psi_{\sigma|\acute{\sigma}}^{p|q}$ in Eq. (6) is symbolized by its associated spin and color: the electron spin and the z–projection of hole total angular momentum are distinguished by blue and red arrows, respectively. Subscript "d" stands for the normalized density of direct matrix element. Right up panel: exchange interaction terms calculated by putting the electron-hole mixed charge densities $\psi_{\uparrow_e}^{e\,*}\psi_{\uparrow_h}^{h}$ or $\psi_{\downarrow_e}^{e\,*}\psi_{\downarrow_h}^{h}$ in the potential, $V_{e_0,h_0}$, formed by the other mixed charge density $\psi_{\uparrow_e}^{e\,*}\psi_{\uparrow_h}^{h}$. Right down panel: exchange interaction terms when the electron-hole mixed charge densities $\psi_{\downarrow_e}^{e\,*}\psi_{\uparrow_h}^{h}$ or $\psi_{\uparrow_e}^{e\,*}\psi_{\downarrow_h}^{h}$ are located inside the potential formed by the other mixed charge density $\psi_{\uparrow_e}^{e\,*}\psi_{\downarrow_h}^{h}$. Subscript "ex" means the normalized density of exchange matrix element, which is 400 times scaled up for the sake of clarity. Only real parts are illustrated for $\langle \uparrow_e \uparrow_h |C| \downarrow_e \downarrow_h \rangle_{ex}$ and $\langle \uparrow_e \downarrow_h |C| \downarrow_e \uparrow_h \rangle_{ex}$.

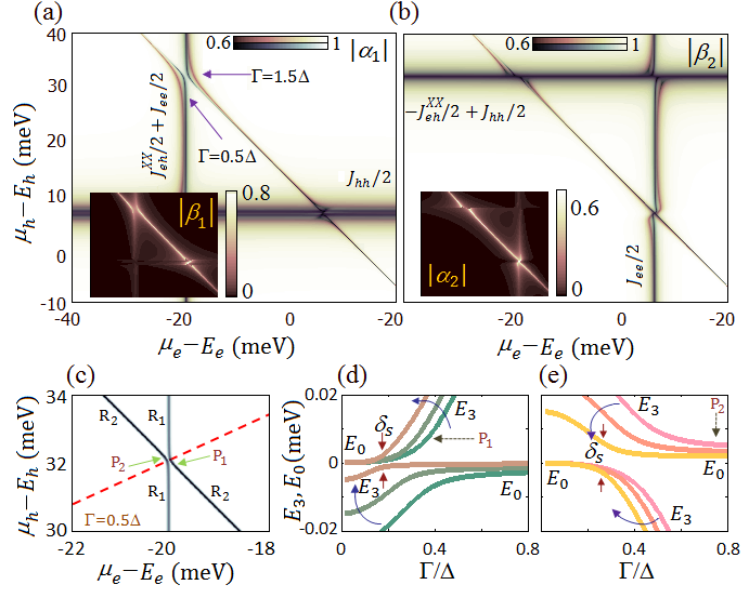

FIG. 3 (a) and (b) $|\alpha_i|$ and $|\beta_i|$ coefficients of the coupled system ground $|G\rangle$ and biexcitonic $|XX\rangle$ states against detunings $\Delta\varepsilon_e$ and $\Delta\varepsilon_h$ for $\Gamma = 0.5\Delta_{Nb}$ and $1.5\Delta_{Nb}$. Sweep window only includes areas near the anticrossing for the sake of clarity. As demonstrated, along the skew region, $|G_e, G_h\rangle$ and $|\Theta_e, \Theta_h\rangle$ components contribute dominantly to $|G\rangle$ and $|XX\rangle$ wavefunctions. (c) $\Gamma = 0.5\Delta_{Al}$, bold blue branch $R_2$ shows the regions where $\delta_s < \Delta_{Al} \sim 0.16$ meV: $|G\rangle$ and $|XX\rangle$ are almost degenerate. Pale blue branch $R_1$ shows the regions where $|G\rangle$ and $|e_1, h_0\rangle = \alpha_3|G_e, G_h\rangle + \beta_3|\Theta_e, G_h\rangle$ become degenerate ($\delta_{|G\rangle,|e_1,h_0\rangle} < \Delta_{Al}$). Dashed red line passes exactly through the anticrossing. $P_1$ and $P_2$ are selected points on the skew branch separated $\sim 200$ µeV from anticrossing. The last two plots illustrate the ground and biexciton energy levels varying versus level broadening. Chemical potential of the electron side $\mu_l^e$ is changed in 5 µeV steps starting from $P_1$ (d) and $P_2$ (e) toward anticrossing while the hole side chemical potential $\mu_l^h$ is kept fixed.

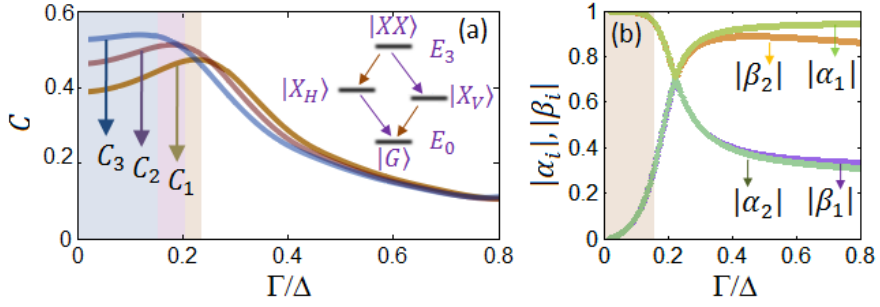

FIG. 4 (a) Concurrence defined in Eq. (9) plotted versus level broadening for $\Delta = \Delta_{Al}$, assuming $\gamma_{|XX\rangle \to |X_V\rangle,|X_H\rangle} = 2\gamma_{|X_V\rangle,|X_H\rangle \to |G\rangle}$ and $\gamma_{|X_V\rangle,|X_H\rangle \to |G\rangle} = 4$ μeV. In the three curves shown here, $C_1$, $C_2$, and $C_3$, chemical potential on the electron side $\mu_l^e$ changes in 2 μeV steps starting from $P_2$ (curve $C_1$) toward anticrossing whereas hole side chemical potential $\mu_l^h$ is held fixed. (b) $|\alpha_i|$ and $|\beta_i|$ coefficients corresponding to curve $C_1$ from (a). The maximum concurrence does not occur exactly where $\delta_s$ is minimized due to the contribution from exciton linewidth $\zeta_X$.